\documentclass{elsarticle}

\usepackage[bookmarks=false]{hyperref}
\hypersetup{pdfstartview={XYZ null null 1.00}}
\usepackage{graphicx}
\usepackage{natbib}
\usepackage{epstopdf}

\newcommand{\be}{\begin{equation}}
\newcommand{\ee}{\end{equation}}
\newcommand{\bes}{\begin{subequations}}
\newcommand{\ees}{\end{subequations}}
\newcommand{\bea}{\begin{eqnarray}}
\newcommand{\eea}{\end{eqnarray}}
\newcommand{\bear}{\begin{equation}\begin{array}}
\newcommand{\eear}[1]{\end{array}\label{#1}\end{equation}}
\def\ba{$$\begin{array}}
\def\ea{\end{array}$$}
\def\bra{$\begin{array}}
 \def\era{\end{array}$}

\begin{document}

\begin{frontmatter}

\title{Has a  Higgs-flavon with a $750$ GeV mass been detected at the LHC13?}

\author{A. Bola\~nos}
\address{Facultad de F\'isica\\ Universidad Veracruzana, 91000, Xalapa, Veracruz, M\'exico}
\author{J.L. Diaz-Cruz, G. Hern\'andez-Tom\'e, G. Tavares-Velasco}
\address{Facultad de Ciencias F\'isico-Matem\'aticas\\
Benem\'erita Universidad Aut\'onoma de Puebla, C.P. 72570,
Puebla, Puebla, M\'exico}

\begin{abstract}
Higgs-flavon fields appear as a part of the Froggatt-Nielsen (FN) mechanism, which attempts to explain the hierarchy of
Yukawa couplings.  We  explore the possibility that the 750 GeV diphoton
resonance recently reported  at the LHC13, could be identified with a low-scale Higgs-flavon field $H_F$ and find the region of the parameter space consistent with CMS and ATLAS data. It is found that the extra vector-like fermions of the ultraviolet completion of the FN mechanism are necessary in order to reproduce the observed signal. We consider a standard model (SM)  extension that contains two Higgs doublets (a standard one and an inert one) and one complex FN singlet. The inert doublet includes a stable neutral boson, which provides a viable dark matter candidate, while
the mixing of the standard doublet and the FN singlet induces flavor violation in the Higgs sector at the tree-level. Constraints on the parameters of the model  are derived from the LHC Higgs data,  which include the
search for the lepton flavor violating decay of the SM  Higgs boson $h\to \bar{\mu}\tau $.
 It is  also found that in some region of the parameter space the model  may give rise to a large branching ratio for the $H_F \to hh$ decay,  of the  order of 0.1, which could be searched for at the LHC.
\end{abstract}

\end{frontmatter}


\section{Introduction}
\label{intro}
The first results of the LHC Run at 13 TeV has shown
surprising hints of a new resonance in the diphoton channel with invariant
mass of 750 GeV \cite{atlas750,cms750}, which clearly represents a signal of physics beyond the standard model (SM). The ATLAS collaboration collected
3.2 fb$^{-1}$ of data and reports a signal with significance of  3.6 $\sigma$ (local),
which becomes  2.3 $\sigma$ (after LEE) \cite{atlas750}, while the CMS collaboration collected an integrated luminosity of
2.6 fb$^{-1}$ and reports a significance  2.6 $\sigma$ (local) that became  2.0 $\sigma$ (after LEE) \cite{cms750}.
As tentative as the signal  could be, it has motivated a large number
of studies that attempt to reproduce its profile (for some works see for instance \cite{750Resonance}).

 Several ideas have been proposed to address the flavor problem \cite{Isidori:2010kg}. For instance,
textures and GUT-inspired relations, flavor symmetries and radiative generation, etc.
The flavor symmetry approach can be supplemented with the Froggatt-Nielsen (FN) mechanism, which
assumes that above some scale $M_F$ there is a symmetry that forbids the  appearance of  Yukawa couplings;
SM fermions are charged under this symmetry [which could be of Abelian type $U(1)_F$].
However, the Yukawa matrices can arise  through  non-renormalizable operators.
The Higgs spectrum of these models could include light Higgs-flavons, which could mix with
the scalar bosons. In these models, the diagonal flavor conserving (FC) couplings of the  SM-like Higgs boson
could deviate from the SM, while FV couplings could be induced at small rates too,
but still  produce detectable signals.
 On the other hand, extending the Higgs sector of the SM opens up the possibility of  including
a scalar dark matter (DM) candidate, such as occurs with the well studied inert doublet model (IDM).
There are important motivations to supplement this model with a complex singlet, for instance to have extra sources
of CP violation, as in the IDM with a complex singlet (IDMS)  recently studied \cite{Bonilla:2014xba}.

In this paper we explore the possibility that the 750 GeV diphoton
resonance \cite{atlas750,cms750}, could be identified with a low-scale Higgs-flavon field $H_F$.\footnote{ A different approach based on a flavor model has appeared very recently
\cite{Bonilla:2016sgx}.}
We work within a SM extension of the IDMS-type  that contains two Higgs doublets and one complex FN singlet.
The mixing of the doublet and the singlet induces FV in the Higgs sector at the tree-level, which can
accommodate the LFV Higgs decay $h \to \bar\mu \tau$ searched for at the LHC \cite{Huitu:2016pwk}.

\section{The model}
\label{model}
We consider a  multi-Higgs  model including one SM-like Higgs doublet $\Phi_s$, an inert-type doublet $\Phi_n$, and
one FN scalar field (SM singlet $S$).
The possibility of having light Higgs-flavon fields was studied in Ref.
\cite{Dorsner:2002wi}, and more recently in \cite{Tsumura:2009yf}.
Besides breaking the EW symmetry,  the $\Phi_s$ doublet gives masses to the quarks and leptons.
By imposing a discrete symmetry, the $\Phi_n$ doublet will be of the inert-type and will contain a DM candidate \cite{Diaz-Cruz:2014pla}.

In our model only the $Z_2$-even fields $\Phi_s$ and $S_F$ acquire vacuum expectation values $v$ and $u$,
respectively. We will use the following field decomposition around the
vacuum state:

\begin{eqnarray}
& \Phi_s = \left( \begin{array}{c} G^+ \\ \frac{1}{\sqrt{2}} \left( v + \phi^0 + i G_z \right)\\
\end{array} \right), \qquad \Phi_{n} = \left( \begin{array}{c} H^+ \\ \frac{1}{\sqrt{2}} \left( H +
i A \right)\\ \end{array} \right), \label{dec_doublets}&\\[2mm]
& S_F = \frac{1}{\sqrt{2}} (u  + s_1 + i p_1 ). \label{dec_singlet}&
\end{eqnarray}


The Higgs potential resembles that of the IDMS model studied in \cite{Bonilla:2014xba}.
To reduce the free parameters, a  $U(1)$ symmetry  identified with the flavor symmetry is invoked. Its breaking helps out to address the
hierarchy of the Yukawa couplings associated with the broad spectra of fermion masses.
The scalar potential is invariant under CP and  takes the  form:

\begin{eqnarray}
V &=& -\frac{1}{2}\left[{m_{1}^2} \Phi^\dagger_s\Phi_s + {m_{2}^2} \Phi_n^\dagger\Phi_n \right]
+ \frac{1}{2}\left[\lambda_1 \left( \Phi^\dagger_s\Phi_s \right)^2
+ \lambda_2 \left(\Phi_n^\dagger\Phi_n \right)^2\right]\nonumber\\
&+&  \lambda_3 \left( \Phi^\dagger_s\Phi_s \right) \left(\Phi_n^\dagger\Phi_n\right) +
\lambda_4 \left( \Phi^\dagger_s\Phi_n \right) \left(\Phi_n^\dagger\Phi_s \right)
+\frac{\lambda_5}{2}\left[\left(\Phi^\dagger_s\Phi_n\right)^2
+\left(\Phi_n^\dagger\Phi_s \right)^2\right]\nonumber \\
&-&\frac{m_3^2}{2} S^*_F S_F -\frac{m_4^2}{2} (S^{*2}_F + S^2_F) + \lambda_{s1} (S^*_F S_F)^2
+ \lambda_{ss} (\Phi^\dagger_s\Phi_s)(S^*_F S_F)\nonumber\\ &+& \lambda_{sn} (\Phi^\dagger_n\Phi_n)(S^*_F S_F)
\label{potIDM1S}
\end{eqnarray}
We are   left with  $U(1)$-symmetric terms
($m_{1}^2$, $m_{2}^2$, $m_{3}^2$, $\lambda_{1-5}$, $\lambda_{s1}$, $\lambda_{ss}$, and   $\lambda_{sn}$) and a $U(1)$-soft-breaking
term ($m_{4}^2$).
An extensive analysis of the CP-violating version of this potential was presented in Ref. \cite{Bonilla:2014xba}. For the CP conserving case, imposing the minimization conditions for $V$ results in the following relations:

\begin{eqnarray}
 m_{1}^2  &=&  v^2 \lambda_1 + u^2 \lambda_{ss} \\
 m_{3}^2 &=& -2 m^2_4 + 2 u^2 \lambda_{s1} + v^2 \lambda_{ss}
\end{eqnarray}
Since we are considering a CP-invariant potential, the CP-even (real)  and CP-odd (imaginary)
components of the mass matrix do not mix. Thus, the mass matrix for the real components in the basis $(\phi^0,H, s_1)$ is given by:

\begin{equation}
M^2_S =
 \left( \begin{array}{ccc}
   \lambda_1 v    &  0    &  \lambda_{ss} u v \\
        0          &   \frac{1}{2} (-m^2_3+\lambda^{+} v^2+ \lambda_{sn} u^2 )   & 0  \\
\lambda_{ss} u v    &   0                & 2\lambda_{s1} u^2
\end{array} \right)
\end{equation}
where $\lambda^{+}= \lambda_3 + \lambda_4  + \lambda_5$.

On the other hand, the mass matrix for imaginary components, in the basis $(G_z,A,p_1)$, reads:

\begin{equation}
M^2_P =
 \left( \begin{array}{ccc}
   0       &  0    &  0 \\
   0       &   \frac{1}{2} (-m^2_3+\lambda^{-} v^2+ \lambda_{sn} u^2 )   & 0  \\
   0        &   0                & 2m^2_3
\end{array} \right)
\end{equation}
where: $\lambda^{-}= \lambda_3 + \lambda_4  - \lambda_5$.

The neutral state $H (A)$ arising from the inert doublet
does not mix with $\phi^0$ nor  $s_1$  $(p_1)$, and the only massive charged state comes from the dark doublet, with
$M^2_{H^+} = \frac{1}{2} (-m^2_3+\lambda_3 v^2+ \lambda_{sn} u^2 ) $.

The real components of the Higgs $(\phi^0)$ and Higgs-flavon fields $(s_1)$ do mix, and
the mass eigenstates are obtained through the standard $2\times 2$ rotation:

\begin{eqnarray}
 \phi^0   &=& \, \, \, \, \cos  \alpha h + \sin  \alpha  H_F \nonumber \\
  s_1    &=& -\sin   \alpha h + \cos \alpha H_F
\end{eqnarray}

The mass eigenstates are $h$, which corresponds to the SM-like
Higgs boson, with $m_h=125$ GeV, whereas  $H_F$ and $A_F$ are the heaviest states. The properties
of $H_F$ will depend on those of the  $h$ Higgs boson  due to their mixing.  $A_F$  does not couple
to the gauge bosons, but it does to the SM fermions, including both FC and FV interactions.

In the forthcoming analysis of the Higgs decays we will use the trilinear vertex $H_F hh$, which is given by:

\begin{eqnarray}
 g_{H_F hh}  &=&  \frac{1}{2} \left[  \lambda_{ss}  (u \cos^3 \alpha  + v \sin^3 \alpha )
 + 2 u \sin^2 \alpha \cos \alpha (3\lambda_{s1} -\lambda_{ss})\right.\nonumber\\
 &+&\left. v \sin \alpha \cos^2 \alpha (3\lambda_1 - 2\lambda_{ss}) \right]\simeq \frac{1}{2}\lambda_{ss} u\equiv \lambda u
\end{eqnarray}


The FN Lagrangian, which  includes the terms that  become  Yukawa
couplings after the $U(1)$ flavor symmetry is spontaneously broken,  is given by:
\begin{equation}
 {\cal{L}}_Y =    \rho^u_{ij}  \left( \frac{ S_F }{\Lambda_F} \right)^{n_{ij}} \bar{Q}_i d_j  \tilde{\Phi}
                + \rho^d_{ij}  \left(\frac{ S_F }{\Lambda_F}\right)^{p_{ij}} \bar{Q}_i u_j \Phi
                + \rho^l_{ij}  \left(\frac{ S_F }{\Lambda_F}\right)^{q_{ij}} \bar{L}_i l_j \Phi  +{\rm H.c.}
\end{equation}
where $n_{ij}$, $p_{ij}$, and $q_{ij}$ denote the combination of Abelian charges for each fermion type.
The Higgs-flavon field $S_F$ is assumed to have flavor charge equal to -1, such that
${\cal{L}}_{Y}$ is $U(1)_F$-invariant.
Then,  the Yukawa couplings arise after the spontaneous breaking  of the flavor symmetry,
i.e. $\lambda_x \sim (\frac{<S_F>}{\Lambda_F})^{n_x}$, where $<S>$ denotes the
Higgs-flavon vacuum expectation  value, while $\Lambda_F$ denotes the heavy mass
scale, which represents the mass of heavy fields that
transmit such symmetry breaking to the quarks and leptons.
For  specific structures of Yukawa matrices for each fermion type,
see \cite{Diaz-Cruz:2014pla}.

In the mass eigenstate basis we have the following Lagrangian for
the Higgs-fermion couplings

\begin{eqnarray}
 {\cal{L}}_Y &=&   \frac{1 } {  v} [ \bar{U} M_u U  +  \bar{D} M_d  D
             +  \bar{L} M_l L ] ( c_{\alpha} h + s_{\alpha} H_F) \nonumber \\
            & & +  \frac{v } { \sqrt{2} u}  [\bar{U}_i \tilde{Z}^u U_j  +  \bar{D}_i \tilde{Z}^d D_j
                +  \bar{L}_i\tilde{Z}^l  L_j ]  ( -s_{\alpha} h + c_{\alpha} H_F + i A_F)
\end{eqnarray}
Here, the information about the size of FV Higgs couplings is contained in the $\tilde{Z}^f$ matrices.
Thus, the (diagonal and non-diagonal) interactions of the scalar bosons ($h,H_F, A_F)$ to the fermions $f_i$
are:

\begin{eqnarray}
(\bar{f}_i f_i h) &=& \frac{c_{\alpha} }{v} \bar{M}^f_{ii} - \frac{s_{\alpha} v }{ \sqrt{2} u} {\tilde{Z}^f}_{ii}  \nonumber \\
(\bar{f}_i f_j h) &=& - \frac{s_{\alpha} v }{ \sqrt{2} u} {\tilde{Z}^f}_{ij}  \nonumber \\
(\bar{f}_i f_i H_F) &=& \frac{s_\alpha }{v} \bar{M^f}_{ii} + \frac{c_\alpha v }{ \sqrt{2} u} {\tilde{Z}^f}_{ii}  \nonumber \\
(\bar{f}_i f_j H_F) & = &  \frac{c_{\alpha} v }{ \sqrt{2} u} {\tilde{Z}^f}_{ij}  \nonumber \\
   ({\bar{f}}_i f_j A_F) & = & \frac{ i v }{ \sqrt{2} u } {\tilde{Z}^f}_{ij}\gamma^5
\end{eqnarray}

Besides the Yukawa couplings, we also need to specify the Higgs couplings to the vector bosons,
which we  write as $g_{h_i VV} =  \chi^{h_i}_V g^{\rm SM}_{hVV} $, with the factor $\chi^{h_i}_V$
given as: $ \chi^h_V = \cos\alpha$ and $ \chi^{H_F}_V = \sin \alpha$.

Moreover, since the Higgs couplings to first generation fermions are highly suppressed,
in order to study the FV Higgs coupling, which depends on the $\tilde{Z}^f$ matrices,
we will consider a 2nd-3rd family sub-system. Namely, for up quarks the $\tilde{Z}^u$ matrix
(in mass eigenstate basis),  is given by:

\begin{equation}
\tilde{Z}^{u}=
 \left( \begin{array}{cc}
  Y^u_{22}      &    Y^u_{23}   \\
  Y^u_{23}      &    2 s_u Y^u_{23}
\end{array} \right)
\end{equation}
and similarly for down quarks and leptons. We find a relation among
the parameters, such that we can express the $\rho^{u,d}_{ij}$ coefficients in terms of  ratios of masses
and the CKM angle $V_{cb} \simeq s_{23}$. Namely, we define:  $r_u= m_c/m_t$, $r_d= m_s/m_b$, $r^u_1= Y^u_{22}/Y^u_{33}$, and $r^u_2= Y^u_{23}/Y^u_{33}$. Similarly $r^d_1= Y^d_{22}/Y^d_{33}$ and $r^d_2= Y^d_{23}/Y^d_{33}$.
Within this approximation we have:
$\tilde{Y}^f_{33} \simeq Y^f_{33}$ for $f=u,d$.
Then,
$r^f_1= r_f+ r^f_2$, and the ratios of Yukawa couplings must satisfy the following relation:
\begin{equation}
r^u_2= r^d_2 \frac{1+r_d}{1+r_u} - \frac{s_{23}}{1+r_u}
\end{equation}

Finally, it is worth  mentioning that the FN mechanism can be ultraviolet (UV) completed via the introduction of heavy mirror fermions.
The exact content depends on the specific model, as well as the desired Yukawa matrix. In general, one needs to introduce vector-like quarks,
with the quantum numbers shown in Table \ref{QuantumNumbers}. As will be
discussed below, extra heavy vector-like quarks are necessary  to reproduce the signal for the 750 GeV resonance
hinted at the LHC.

\begin{center}
\begin{table}
\begin{center}
\begin{tabular}{| c | c | c | c | c |}
\hline
 Type  & $SU(3)_c$ & $SU(2)_L$     &  $U(1)_Y$ & $N_f$  \\
\hline
  $P \, (\bar{P})$ & 3  & 2  & $\frac{1}{6}$  & 3 (3)  \\
 \hline
 $U \,  (\bar{U})$ & 3  & 1 & $\frac{2}{3}$   & 3 (3)  \\
  \hline
  $D \, (\bar{D})$ & 3  & 1 & -$\frac{1}{3}$  & 3 (3)   \\
\hline
\end{tabular}
\end{center}
\caption{Heavy quarks in the minimal FN mechanism.
\label{QuantumNumbers}}
\end{table}
\end{center}

\section{Scenarios for  the Higgs-flavon  couplings}
\label{constraints}

We now  turn to discuss the formula necessary to express all the relevant Higgs-flavon couplings, which
will allow us to define some benchmark points:

 {\bf{1.  FC Higgs couplings.}} Firstly, we will use LHC data to derive bounds on the Higgs-flavon couplings,
following the analysis presented in Ref. \cite{Giardino:2013bma}. The deviation from the SM Higgs couplings are assured to be small and
are expressed as: $g_{hXX}= g^{sm}_{hXX} (1 + \epsilon_X)$. The results obtained in \cite{Giardino:2013bma} give
the following allowed ranges with 95 \% C.L.:
$\epsilon_t= -0.21\pm 0.23$, $\epsilon_b= -0.19\pm 0.3$, and $\epsilon_{\tau}= 0 \pm 0.18$;
while for the $W$ and $Z$ gauge bosons it is found $\epsilon_W= -0.15\pm 0.14$ and $\epsilon_Z= -0.01\pm 0.13$.

We will use the strongest constraints, which come from a combination of $\epsilon_Z$ and
$\epsilon_t$, in such a way that the resulting constraints on the mixing  angles is
$0.86 < \cos\alpha <1.0$.

 {\bf{2. FV Higgs couplings to up-type quarks.}} As far as the couplings with up-type quarks are concerned, we will
follow the method outlined in Ref. \cite{Diaz-Cruz:2014pla}. Namely, we will consider the following sample values: $r^2_d= 0.05$, $0.1$, and $0.3$.
Table \ref{Zmatrix} shows the values of the entries for the up-type quark $\tilde{Z}^u$ matrix in the 2nd-3rd family scenario.
We choose to focus  on the up-quark sector, because we want to obtain an estimate for the most relevant
predictions of the model.

 {\bf{3. LFV Higgs couplings.}} These couplings are written in terms of the parameters $\rho_{ij}$, which appear
 in the charged lepton mass matrix, and are of the order of $O(1)$.  Namely,
\begin{eqnarray}
 \tilde{Z}^l_{33} &=& 2\sqrt{2} \frac{m_{\tau} }{v} \simeq 1.95 \times 10^{-2} \nonumber \\
  \tilde{Z}^l_{23} &=& 4 \lambda^4 \rho^l_{23} \simeq  10^{-2} \rho^l_{23} \nonumber \\
 \tilde{Z}^l_{22} &=&  4.1\times 10^{-2} \rho^l_{23} + 2.41 \times 10^{-3}
\end{eqnarray}

We will consider values of $\rho^l_{23}=0.25, 0.75$.


\begin{center}
\begin{table}
\begin{center}
\begin{tabular}{| c| c | c | c |}
\hline
 Scenario  & ${\tilde{Z}}^u_{33}$ & ${\tilde{Z}}^u_{23}$ & ${\tilde{Z}}^u_{22}$  \\
 \hline
   X1     & $4 \times 10^{-4}$   & $2 \times 10^{-2}$    & $2 \times 10^{-4}$  \\
 \hline
   X2     &  $1.4 \times 10^{-2}$   & $1.2 \times 10^{-1}$    & $7.2 \times 10^{-3}$  \\
 \hline
   X3     & 0.27   & 0.52    & 0.14  \\
 \hline
\end{tabular}
\end{center}
\caption{ Relevant elements of the matrix $\tilde{Z}^u_{ij}$ for up-type quarks.
\label{Zmatrix}}
\end{table}
\end{center}

\begin{figure}[!hbt]
\centering
\includegraphics[width=4in]{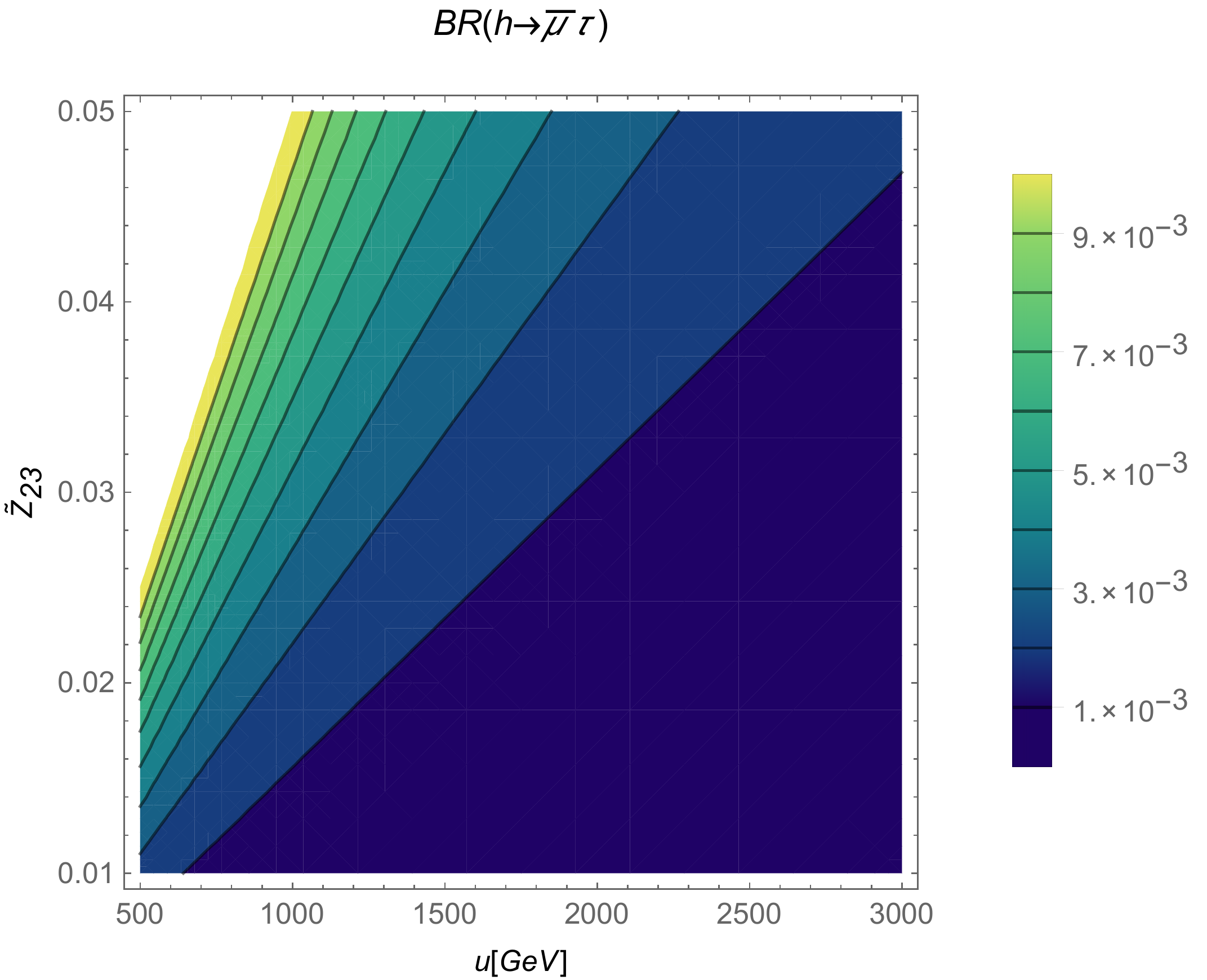}
\caption{Contour plot for the branching ratio of the flavor violating decay of the SM Higgs boson $h\to \bar{\mu}\tau$ in the $u-\tilde{Z}_{23}$ plane.    For the parameter values we use $s_\alpha=0.4$ and $\tilde{Z}_{33}=0.15$.  \label{BRmutau} }
\end{figure}

An interesting probe of FV Higgs couplings is provided by the decay $h\to \bar{\mu}\tau$,
which was initially studied in Refs. \cite{Pilaftsis:1992st, DiazCruz:1999xe}. Subsequent studies on the
detectability of the signal appeared soon after \cite{Han:2000jz, Assamagan:2002kf, Kanemura:2005hr}.
Precise loop calculations with massive neutrinos, SUSY and other models were worked out in
\cite{DiazCruz:2002er,Arganda:2004bz,BRignole:2004ah,DiazCruz:2008ry}.
A search for this decay at the LHC Run I \cite{Khachatryan:2015kon} observed a slight excess of signal events with a significance of 2.4 standard deviations.
Several works appeared trying to explain that result \cite{Vicente:2015cka}. However a  recent report \cite{CMS:2016qvi} rules out any  excess and sets the limit $BR(h\to \bar{\mu}\tau) < 1.2 \times 10^{-2}$  with 95\% C.L. Such a bound is  very loose and
irrespectively of the dismissal of the excess, the search for LFV Higgs decays represents a great opportunity to find new physics at the LHC Run II. We show in Figure \ref{BRmutau} the contour plot for the branching ratio of the SM Higgs boson decay  $h\to \bar{\mu}\tau$ in the $u-\tilde{Z}_{23}$ plane, with $s_\alpha=0.4$ and $\tilde{Z}_{33}=0.15$. We observe that values as large as $10^{-2}$ can be reached for $u$ around 500 GeV and $\tilde{Z}_{23}=0.02$. An improvement of the experimental limit on the $h\to \bar{\mu} \tau$ would put strong constraints on the parameter values.

\section{Higgs-flavon decay modes}
\label{flavondecays}

\begin{figure}[!hbt]
\centering
\includegraphics[width=4.5in]{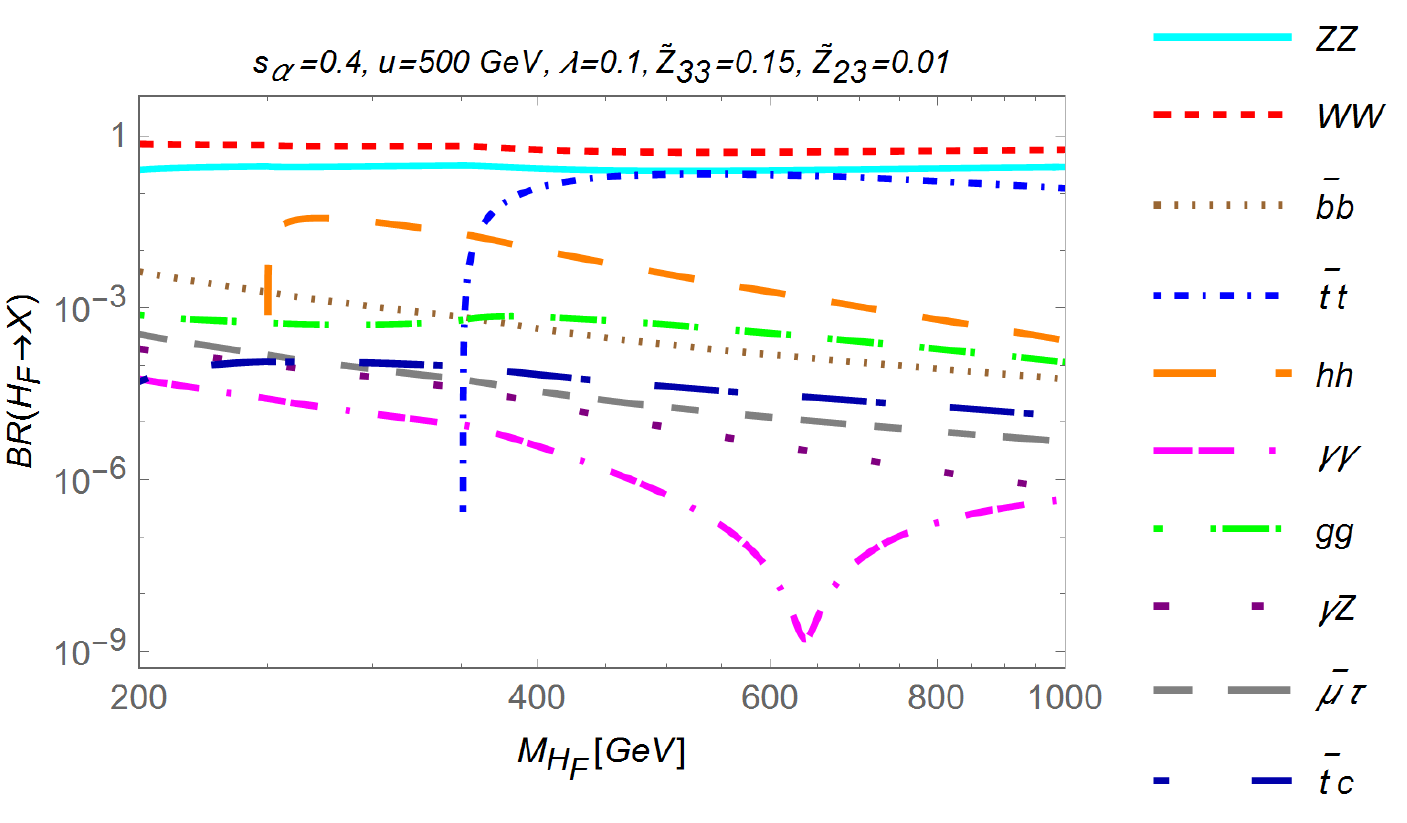}
\caption{Branching ratios  for the relevant  decays of the  Higgs-flavon state $H_F$ as functions of $M_{H_F}$ for the indicated  set of parameter values.
\label{HFBRsa1}}
\end{figure}

The calculation of the two-body tree-level decays is straightforward and so is that of the two-body one-loop decays. To illustrate the behavior of the Higgs-flavon decays we will consider two scenarios of interest, namely, a middle and a tiny value of the mixing angle $s_\alpha$. We  first consider the following set of parameter values: $s_\alpha=0.4$, $u=500$ GeV, $\lambda=0.1$,  $\tilde Z_{33}=0.15$,  and $\tilde Z_{23}=0.01$. The relevant  branching ratios of the Higgs-flavon decays, as functions of the Higgs-flavon mass $M_{H_F}$, are shown in Figure \ref{HFBRsa1}. We observe that in this scenario  the  decay modes  $H_F\to WW$, and $H_F\to ZZ$ are the dominant ones, with branching ratios of the order of 0.7 and 0.35, respectively. These decay channels remain the dominant ones even after the threshold of the $H_F\to t\bar{t}$ decay, which in turn can reach a branching ratio of around $0.2$ at most. We also notice that the decay $H\to hh$ could reach a branching ratio of the order of  ten percent approximately, although such a value   is highly dependent on the value of the $\lambda$ parameter.  Such a branching ratio would open up the possibility for the  search of this decay mode at  LHC13. As far as the one-loop induced decays are concerned, they are very suppressed. In particular, the branching ratio of the $H_F\to \gamma\gamma$ decay shows a large dip around 600 GeV, where it is negligible. For the parameter values used, the flavor changing decays  $H_F\to \bar{\mu}\tau$ and  $H_F\to \bar{c}t$ can reach branching ratios of the order of $10^{-3}-10^{-4}$ for  a Higgs flavon with intermediate mass.

\begin{figure}[!hbt]
\centering
\includegraphics[width=4.5in]{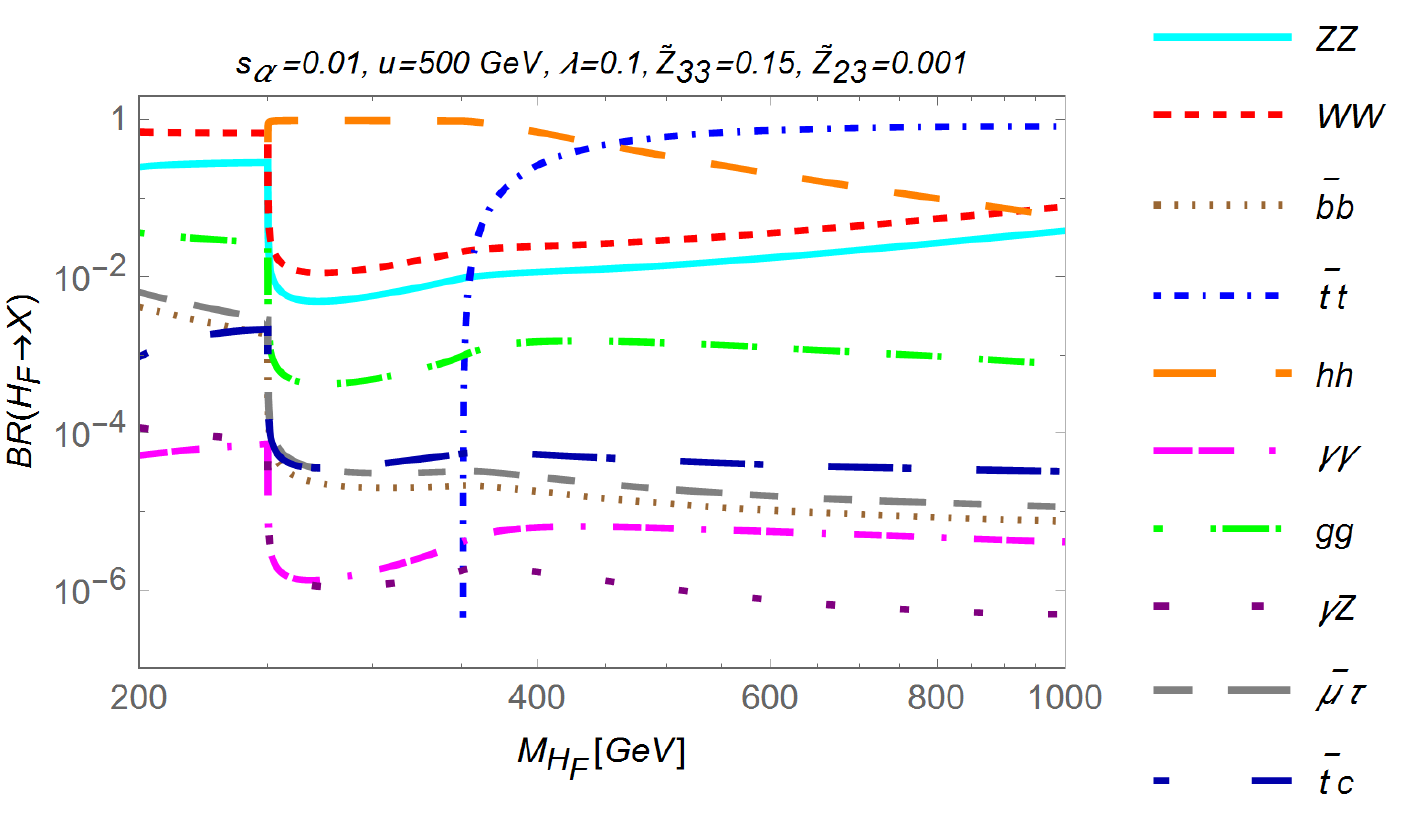}
\caption{Branching ratios  for the relevant  decays of the Higgs-flavon state $H_F$ as functions of $M_{H_F}$ for the indicated  set of parameter values.
\label{HFBRsa2}}
\end{figure}

We now turn to analyze the  scenario with very small  $s_\alpha$, namely, we take  $s_\alpha=0.01$ and use  the same values as above  for all the remaining parameters. The branching ratios for the decays  of the Higgs-Flavon change  are shown  in Figure \ref{HFBRsa2}   as functions of the Higgs-flavon mass.  As expected due to the  dependence of the  Higgs-flavon couplings    on $s_\alpha$ , there is a notorious change in the behavior of the decay widths. For an intermediate mass, the dominant decay channels are still those into a gauge boson pair, but the $H_F\to hh$ decay becomes the dominant one after it is open and until the threshold of the $H_F\to \bar{t}t$ decay is reached. For a heavier Higgs-flavon, the $H_F\to\bar{t}t$ becomes dominant, which is due to the extra term proportional to $c_\alpha$  appearing in the associated coupling constant. The one-loop induced decays have an enhanced branching fraction, but the $H_F\to\gamma\gamma$ and $H_F\to Z\gamma$ channels are still very suppressed. For a lower value of $s_\alpha$  the $H_F\to WW$ and $H_F\to ZZ$ decay widths, though non-vanishing, will become considerably suppressed, which means that the $H_F\to gg$ and $H_F\to \bar{t}t$ channels would become the dominant ones, with the decay $H_F\to\gamma\gamma$ having an enhanced branching ratio. All other Higgs-flavon decays to light fermions would have a negligibly branching ratio. We will analyze below the scenario in which extra vector-like fermions from the UV completion of the Higgs-flavon model contribute at the one-loop level to the $H_F\to gg$ and $H_F\to \gamma\gamma$ decays. This could result in a significant change in the behavior of the Higgs-flavon decays.

\begin{figure}[!hbt]
\centering
\includegraphics[width=4in]{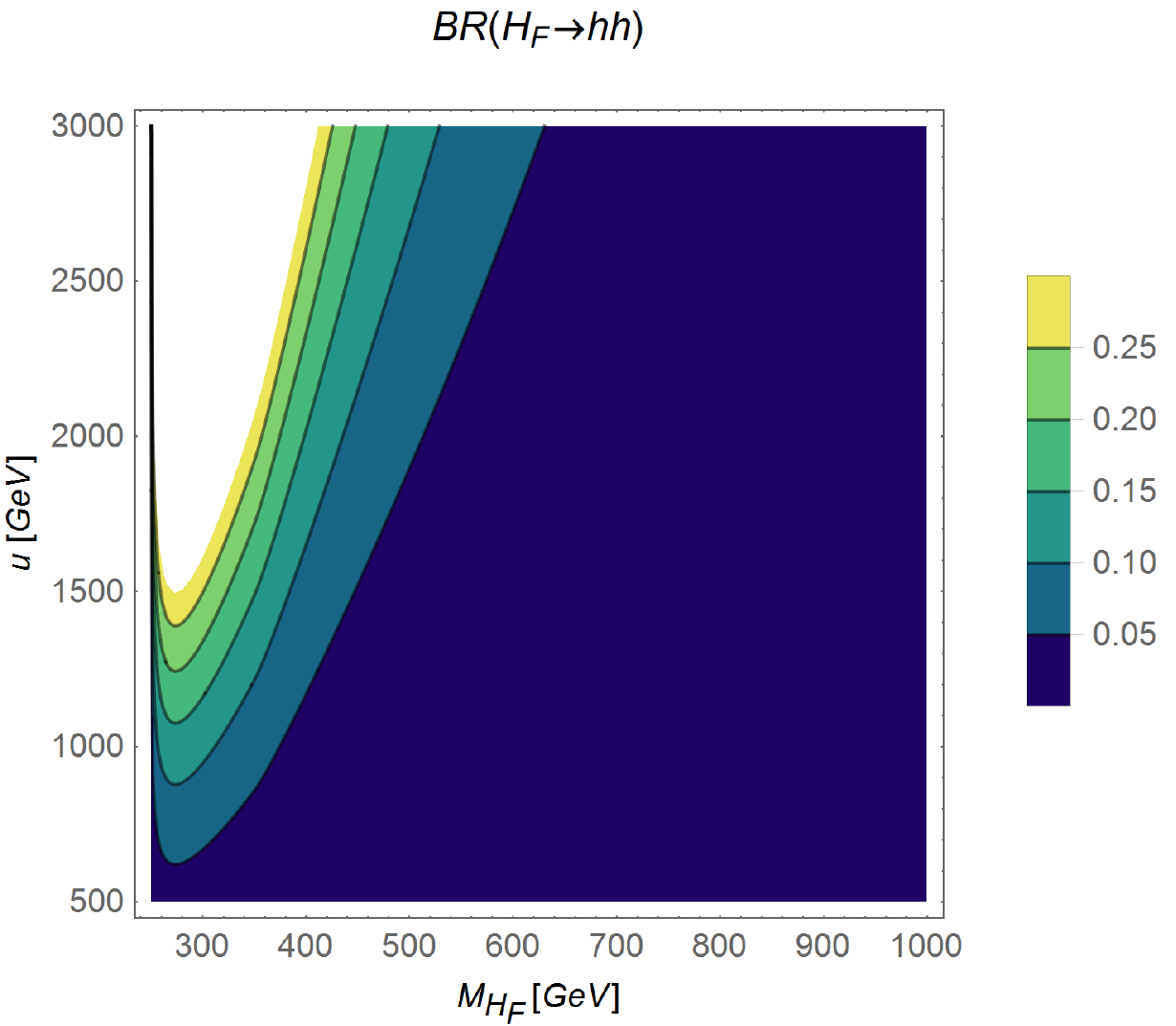}
\caption{Contour plot for the $H_F\to hh$ branching ratio in the $u-M_{H_F}$ plane. For the parameter values we use $s_\alpha=0.4$, $\lambda=0.1$, $\tilde{Z}_{33}=0.15$ and $\tilde{Z}_{23}=0.01$.
\label{HFtohhdecay}}
\end{figure}

It is worth analyzing more detailed the  $H_F\to hh$ decay. Figure \ref{HFtohhdecay} shows that in  the region  of the parameter space enclosed by 500 GeV$ <u < 3000$ GeV and 250 GeV $< M_{H_F} < 1000$ GeV, the branching fraction $BR(H_F\to hh)$ can be of the order of about 0.01, which seems amenable to be searched for at the LHC13.

\section{ The Higgs-flavon as the 750 GeV diphoton resonance}
\label{resonance}

Besides extending the limits on new physics scale, the ATLAS and CMS collaborations have reported
preliminary evidences for a new resonance at 750 GeV in the two-photon channel, which could come
from either a new spin-0 or spin-2   particle.  We will explore the possibility that such a resonance could be identified with the Higgs-flavon
 field $H_F$. By studying the decays of this Higgs-flavon, we can identify the regions of the parameter space that would accommodate  the new 750 GeV signal. A summary of the profile of diphoton resonance at 750 GeV,
shows the following:


 {\bf 1.} ATLAS excess of about 14 events (with selection efficiency 0.4) appears in
 at least two energy bins,  suggesting a width of about 45 GeV
(i.e. $\Gamma/M \simeq 0.06$).  The best CMS fit has a narrow width, while
 assuming a large width  ( $\Gamma/M \simeq 0.06$) decreases the significance,
which corresponds to a  cross section of about 6 fb.

 {\bf 2.} The anomalous events are not accompanied by significant missing energy,
 nor leptons or jets. No resonances at invariant mass 750 GeV are seen in the
 new data in $ZZ$, $W^+ W^-$, or $jj$ events, and  no $\gamma\gamma$ resonances were seen in Run 1 data at $\sqrt{s} = 8$ TeV, although both CMS and ATLAS  data showed a mild upward fluctuation at $m_{\gamma\gamma}= 750$ GeV.
 The data at $\sqrt{s} = 8$ TeV and 13 TeV are compatible at 2 $\sigma$ if the signal
 cross section grows by at least a factor of 5.

 {\bf 3.} For a spin-0 resonance produced from gluon fusion and
decaying mainly into two photons,  the signal rate is reproduced for
\begin{equation}
R^g_{\Gamma} = \frac{\Gamma_{\gamma\gamma} \Gamma_{gg} }{ MM} \simeq 1.1 \times 10^{-6}  \frac{\Gamma}{ M}\simeq 6\times 10^{-8},
\label{Rg}
\end{equation}
 with $M$ the scalar boson mass.

{\bf 4.}When  the resonance $S$ is produced from bottom
quark annihilation, the signal is reproduced for

\begin{equation}
R^b_{\Gamma} = \frac{\Gamma_{\gamma\gamma} \Gamma_{bb} }{ MM} \simeq 1.9 \times 10^{- 4}  \frac{\Gamma}{ M}\simeq 1.1 \times 10^{-5}.
\label{Rb}
\end{equation}

{\bf 5.} The combined data from ATLAS and CMS at $\sqrt{s}=$ 8 and $\sqrt{s}=13$ TeV result in the following production cross section for the diphoton channel

\begin{equation}
\sigma(pp\to S\to \gamma\gamma)=6.6\pm 1.3 \quad\rm{fb}.
\end{equation}

\begin{figure}[!hbt]
\centering
\includegraphics[width=2.5in]{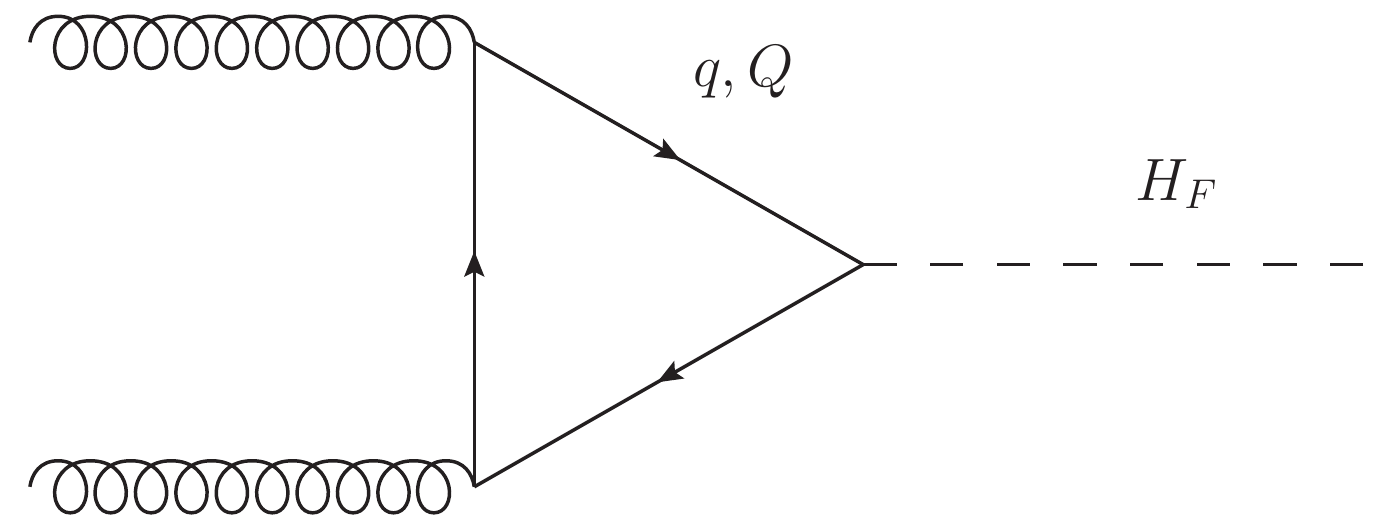}
\caption{Feynman diagram for the leading order contribution to Higgs-flavon production via gluon fusion. In the loop can circulate a SM quark or a new vector like quark of the UV completion of the Higgs-flavon model. A similar diagram induces the $H_F\to \gamma\gamma$ decay except that we need to include  contributions from all other electrically charged particles.
\label{GluonFusion}}
\end{figure}

At the LHC the Higgs-flavon would be mainly produced via gluon fusion mediated by the triangle diagram carrying SM quarks, as shown in Figure \ref{GluonFusion}. The dominant contribution would arise from the top quark. We will consider that there are also  contributions coming from the  heavy vector-like quarks predicted by the UV completion of the Higgs-flavon model. Apart from reproducing the experimental data for the diphoton decay width at $\sqrt{s}=13$ TeV,  the Higgs-flavon also must satisfy  the experimental  bounds set by the ATLAS and CMS collaborations (see Table \ref{CMSATLASBounds}) on  the $\sqrt{s}=8$ TeV  cross section for the production of a  scalar resonance decaying into  gauge boson pairs, gluon pairs,  etc. We will thus examine whether there is a region of the parameter space of the Higgs-flavon model that is in accordance with experimental data.
A quick glance at Figures \ref{HFBRsa1} and \ref{HFBRsa2} allows us to conclude that the following conditions  are to be fulfilled: a very small $s_\alpha$ to achieve  small $H_F\to WW$ and $H_F\to ZZ$ branching ratios,  negligible FV couplings in order to suppress the tree-level decays $H_F\to \bar{\mu}\tau$ and  $H_F\to \bar{c}t$, and  an increase of the Higgs-flavon diphoton production.  The latter can only be achieved through an enhancement of the $H_F\to \gamma\gamma$ decay width along with  an increase of the gluon fusion production mode, which requires the introduction of additional loop contributions from charged/colored particles. Since the contribution of a singly charged scalar  is rather suppressed, an enhancement of the $H_F\to \gamma\gamma$ decay can be achieved with  the addition of extra  vector-like fermions, which can also enhance the $H_F \to gg$ partial width, so a $pp\to H_F\to \gamma\gamma$ cross section of the order of 1-10 fb can be reached at $\sqrt{s}=13$ TeV.  Vector-like fermions are required to not to spoil  the constraints on electroweak precision data. Within the context of our model these heavy quanta could be naturally identified with the heavy vector-like fermions that would arise from the UV completion of the FN mechanism.

\begin{table}[!hbt]
\begin{center}
\begin{tabular}{ccc}
\hline
X&CMS bound [fb]&ATLAS bound [fb]\\
\hline
$WW$&220\cite{Khachatryan:2015cwa}&38\cite{Aad:2015agg}\\
\hline
$ZZ$&27\cite{Khachatryan:2015cwa}&12\cite{Aad:2015kna}
\\
\hline
$\bar{t}t$&600\cite{CMS:lhr}&700\cite{Aad:2015fna}
\\
\hline
$hh$&52\cite{Khachatryan:2015yea}&35\cite{Aad:2015xja}\\
\hline
$gg$&$1800$\cite{Khachatryan:2016ecr}&--\\
\hline
$Z\gamma$&--&6\cite{Aad:2014fha}\\
\hline
$\gamma\gamma$&1.3\cite{cms750}&10\cite{atlas750}\\
\hline
\end{tabular}
\end{center}
\caption{Experimental upper limits imposed by the CMS and ATLAS collaborations  on the $pp\to S\to X$ cross section at $\sqrt{s}=8$ TeV with 95 \% C.L. for a scalar resonance $S$ with a  mass of $750$ GeV. \label{CMSATLASBounds}}
\end{table}

A detailed discussion of the FN UV completion is beyond the purpose of this letter and  we refrain the interested reader to \cite{Ellis:2014dza} and References therein. No matter the details of the specific vector-like fermion model, for the purpose of our analysis is enough to consider an scenario with $N$  degenerate  vector-like quarks with the following effective interaction to the Higgs-flavon

\begin{equation}
{\cal L}=i\frac{C_{Q} m_{Q}}{v} \bar{Q} Q  H_F,
\end{equation}
where $C_Q$  stand for the coupling constant that can be known once a specific model is considered.  There are models that also predict vector-like leptons but they do not yield the necessary enhancement to the $pp\to H_F\to \gamma\gamma$   cross section to reproduce the observed   signal and will not be considered here.  As for the couplings of the vector-like quarks  to the SM gauge bosons,
they can be written as:
 \begin{equation}
{\cal L}= e q_Q\bar{Q}\gamma^\mu Q A_\mu+g_s\bar{Q}\gamma^\mu Q G_\mu+\frac{g}{c_W}\bar{Q}\gamma^\mu(T_3-s_W^2 Q ) QZ_\mu.
 \end{equation}
So, the calculation of the fermion loop of Fig. \ref{GluonFusion} proceeds as usual. The result for
the two-photon decay width, including contributions of charged fermions and the $W$ gauge boson can be written as
\begin{equation}
\Gamma(H_F\to \gamma\gamma)=\frac{\alpha^2m^3_{H_F}}{1024\pi^3m_W^2}\left|\sum_{s=f,W^\pm} A_s^{H_F\gamma\gamma}\left(\tau_s\right)\right|^2,
\label{Htogammagamma}
\end{equation}
where $\tau_{s}= 4m_s^2/m_{H_F}^2 $ and
\begin{equation}
A_s^{H_F\gamma\gamma}(x)=\left\{
\begin{array}{cr}
-\sum_f \frac{2m_W g_{H_F \bar{f}f} N_c Q_f^2}{m_f}\left[2x(1+(1-x)f(x))\right]&s=f,\\ \\
 \frac{ g_{H_F WW}}{m_W}\left[2+3x+3x(2-x)f(x)\right]&s=W,
 \end{array}\right.
\end{equation}
with $g_{H_F \bar{f}f}$ and $g_{H_F WW}$ the respective coupling constants and
\begin{equation}
f(x)=\left\{
\begin{array}{cr}
\left[\arcsin\left(\frac{1}{\sqrt{x}}\right)\right]^2&x\ge1,\\
-\frac{1}{4}\left[\log\left(\frac{1+\sqrt{1-x}}{1-\sqrt{1-x}}\right)-i\pi\right]^2&x<1.
\end{array}
\right.
\end{equation}
The contributions of a singly charged scalar boson is subdominant and can be neglected. For a heavy Higgs-flavon, the main contribution arises from the heaviest charged fermion. As for the $H_F\to gg$ decay,  the respective decay width can be obtained from \ref{Htogammagamma} by taking the quark contribution only and making the replacement $NQ_f^2\to \sqrt{2}$ \cite{Djouadi:2005gi}. Next-to-leading order contributions has also been calculated an the results are summarized for instance in Ref.  \cite{Djouadi:2005gi}. The production cross section of a scalar resonance decaying into the $X$ channel  is given by

\begin{equation}
\sigma(pp\to H_F\to X)=\sigma(pp\to H_F)_{GF}BR(H_F\to X),
\end{equation}
where $\sigma(pp\to H_F)_{GF}$ is the cross section for the production of a scalar resonance at the LHC \cite{Djouadi:2005gi}.

\begin{figure}[!hbt]
\includegraphics[height=2.25in]{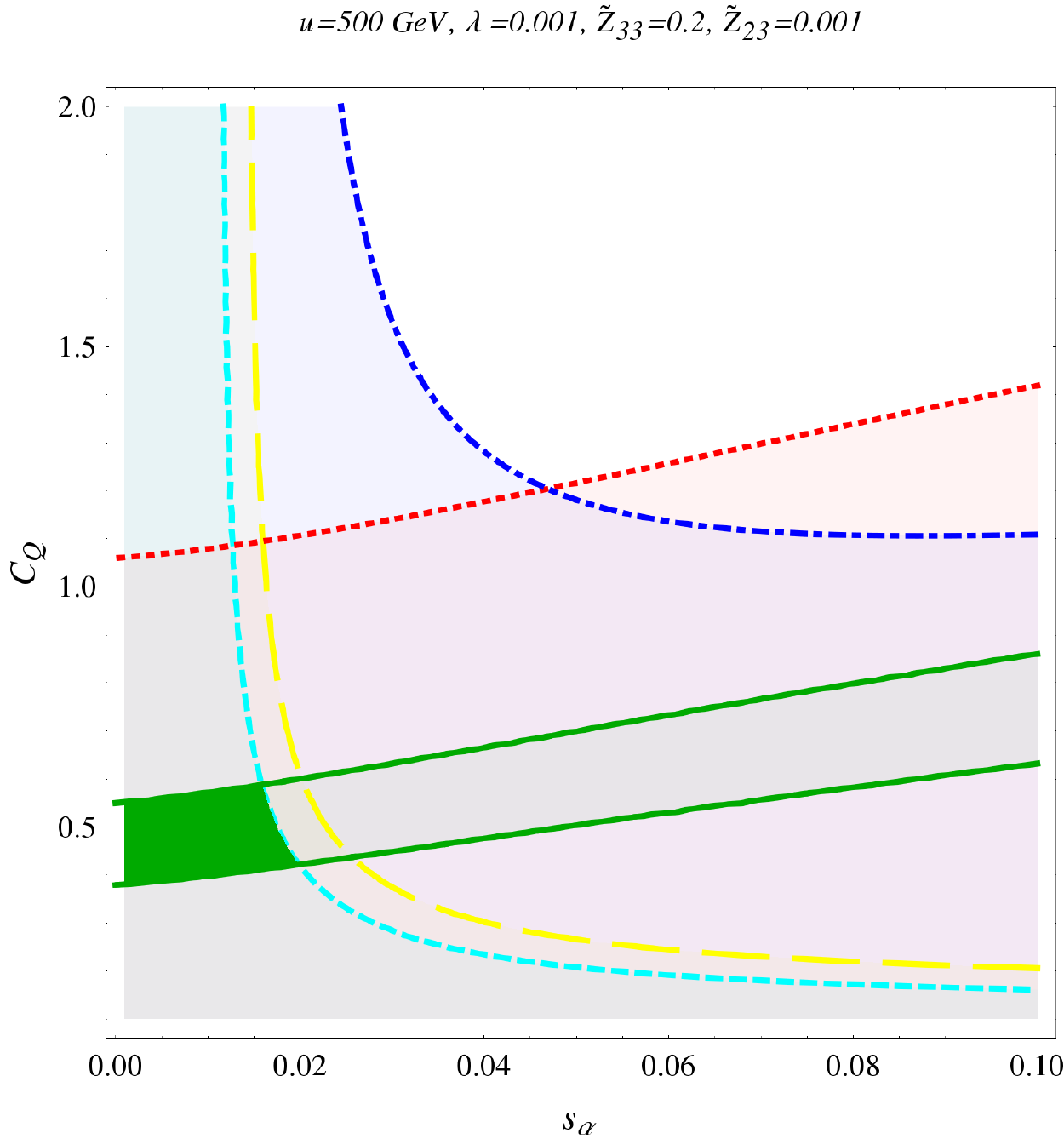}\hspace{2mm}\includegraphics[height=2.25in]{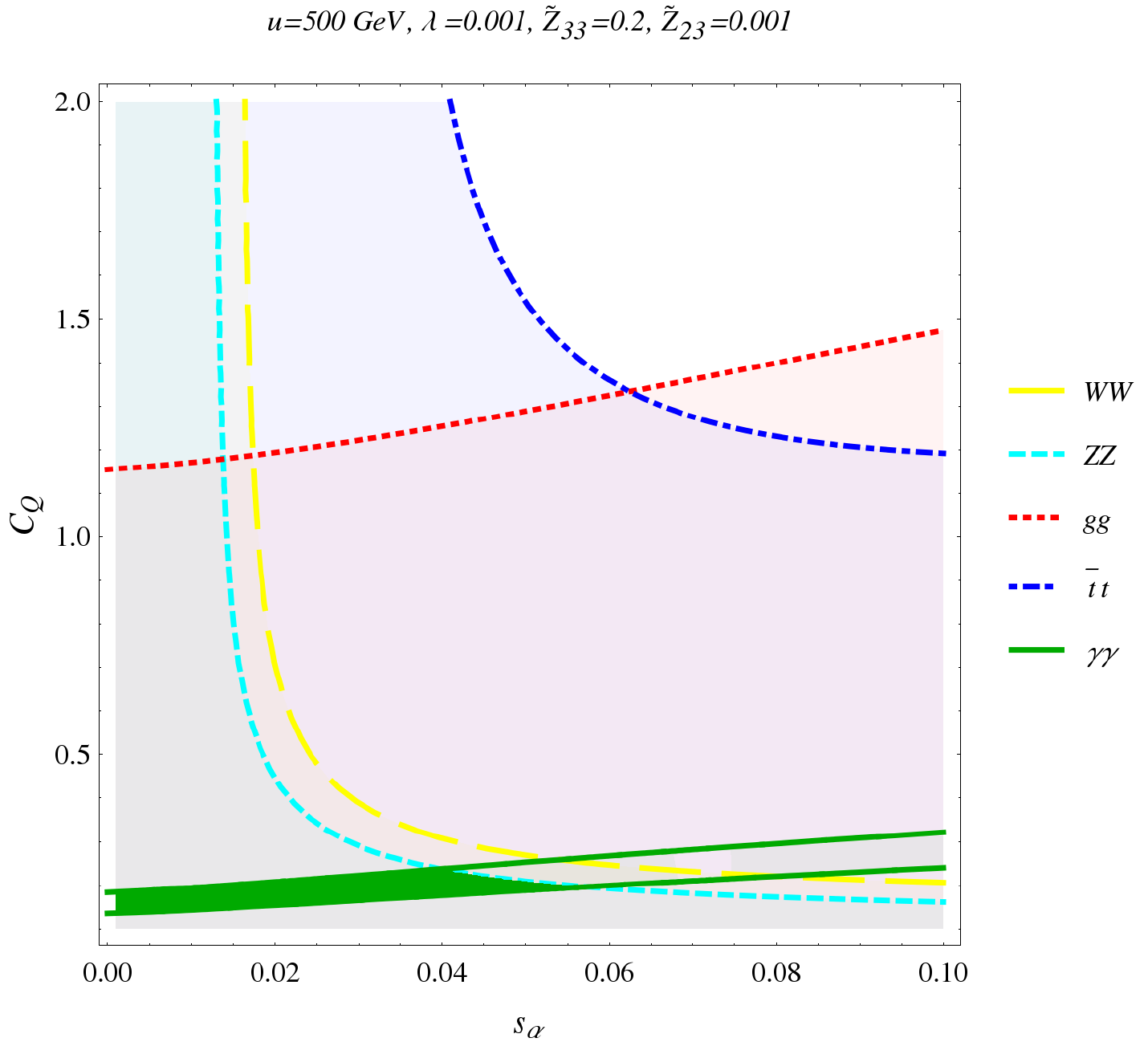}
\caption{Area allowed in the $C_Q$ vs $s_\alpha$ plane by the LHC constraints on  the $\sqrt{s}=8$ TeV $pp\to H_F\to X$  production cross section   mediated by a $750$ GeV   Higgs-flavon for the indicated set of parameter values. We consider  the addition of 3 vector-like  quarks with mass $m_Q=1000$ GeV and charge $2/3$ (left plot) and $5/3$ (right plot). The area below each curve is the one allowed by the particular production mode  at $\sqrt{s}=8$ TeV (see Table \ref{CMSATLASBounds}) and the dark area represents the region where the $\sqrt{s}=13$ TeV $pp\to H_F \to \gamma\gamma$ cross section lies between $6.6\pm 1.3$ fb.
\label{bound1}}
\end{figure}

We will now consider a scenario with 3 degenerate vector-like charge $2/3$ quarks $Q$ (they can be introduced in the model as $SU(2)$ singlets as shown in Table \ref{QuantumNumbers})
and find the region in the $C_Q$ vs $s_\alpha$ plane consistent with the LHC Run I bounds on the production cross section
 of a scalar resonance decaying into a final state $X$, as shown  in Table \ref{CMSATLASBounds},  for
the following set of parameter values: $u=500$ GeV, $\tilde Z_{33}=0.2$,
$\lambda=0.001$, and  $\tilde Z_{23}=0.001$. It means that we are assuming that the
$H_F\to hh$, $H_F\to \bar{\mu}\tau$, and $H_F\to \bar{c}t$
decay channels have a negligible decay width. For the mass of the vector-like quarks we use  $m_Q=1000$ GeV  to fulfill the current experimental bounds. The results are shown in  the left plot of Fig. \ref{bound1}, where the area below each curve is consistent with the LHC Run I data for  the production cross section $\sigma(pp\to H_F\to X)$ and
the dark area is the one in which the diphoton cross section $\sigma(pp\to H_F\to \gamma\gamma)$
lies between $6.3\pm 1.3$ fb, thereby reproducing  the observed diphoton anomaly.  It is important to notice that we choose to use the strongest constraints of Table \ref{CMSATLASBounds}. To estimate the production cross section we implemented a code with the formulas  for the decay widths of a scalar Higgs boson  as well as the gluon fusion cross section \cite{Djouadi:2005gi}  and used the CT10 gluon parton distributions \cite{Dulat:2013kqa}.
We note that the experimental data on the $ZZ$ and $gg$ final
states provides strong constraints on the Higgs-flavon couplings but there is a surviving tiny area
in which the Higgs-flavon model can reproduce the experimental data on the diphoton resonance while still
being consistent with the experimental constraints on the $pp\to S\to X$ production cross section. For $(s_\alpha, C_Q)$ values lying inside the allowed area,  the dominant decay modes are $H_F\to gg$ and $H_F\to \bar{t}t$. If a larger number of vector-like charge $2/3$ quarks are considered, the allowed area will shift downwards and lower $C_Q$ values will be allowed. While a large number of vector-like charge $-1/3$ quarks would be required to reproduce the diphoton signal, there is also the possibility of   vector-like  quarks of exotic charge. We consider an scenario with 3 vector-like charge $5/3$ quarks, which can be introduced in a hypercharge $7/6$ $SU(2)$ doublet and will be accompanied by a vector-like charge $2/3$ quark.  We take $m_Q=1000$ GeV and show the resulting constraints  in the right plot of Fig. \ref{bound1}. In this scenario the allowed region not only has shifted downwards but also has shrunk considerably. It is worth mentioning that the experimental limits on the $\bar{b}b$, $hh$ and $Z\gamma$ final states provide no useful constraints. Also the Higgs-flavon decays into light quarks $H_F\to \bar{q}q$  pose no problem to satisfy the LHC  dijet constraints as the corresponding coupling constants are proportional to $s_\alpha$ and thus yield a negligible decay rate for very small $s_\alpha$. Furthermore, these vector-like quarks would not produce dangerous effects on the loop induced SM Higgs couplings if there is a small mixing of the Higgs-flavon with the SM Higgs boson.

\begin{figure}[!hbt]
\centering
\includegraphics[height=2.25in]{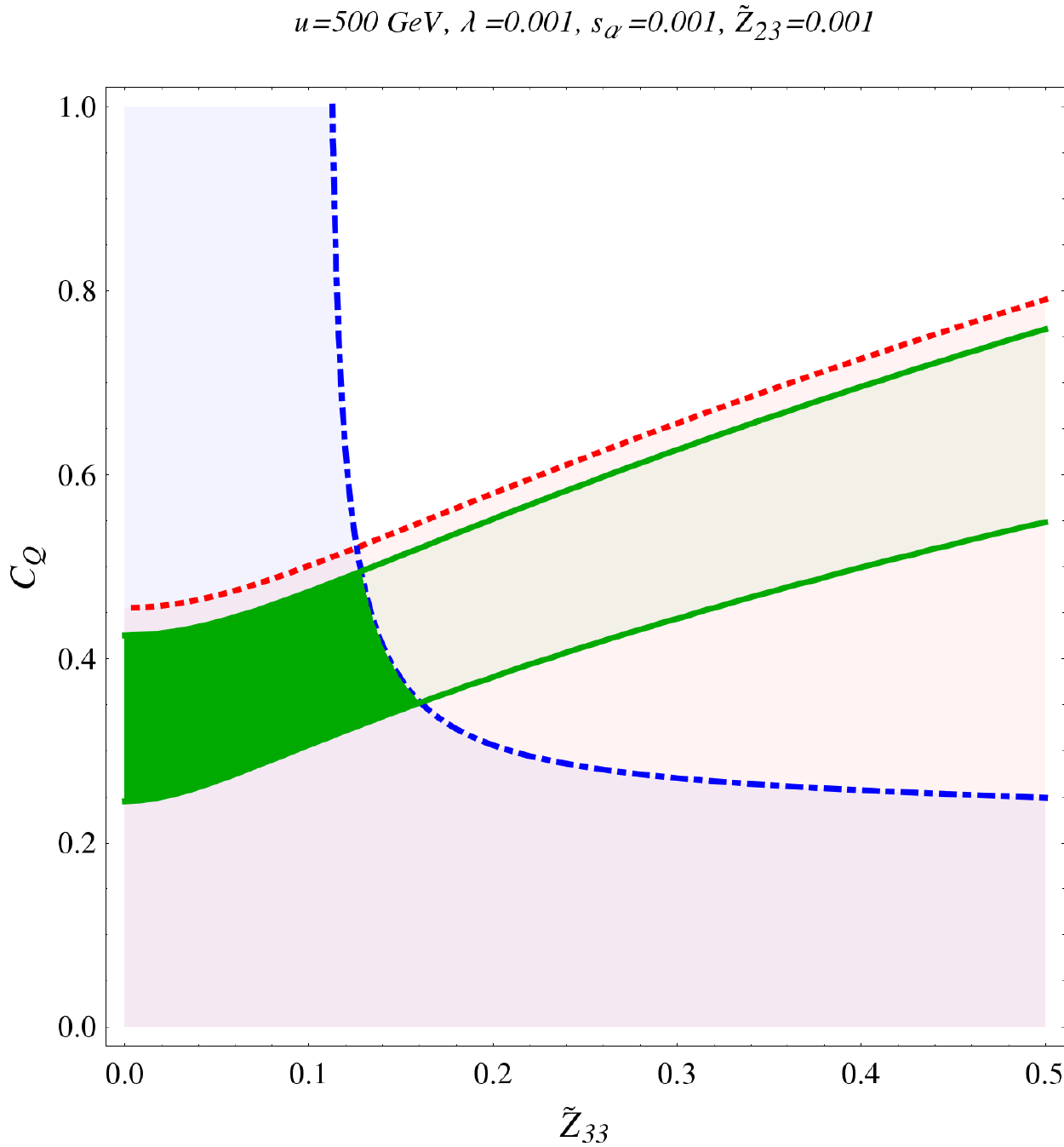}\hspace{2mm}\includegraphics[height=2.25in]{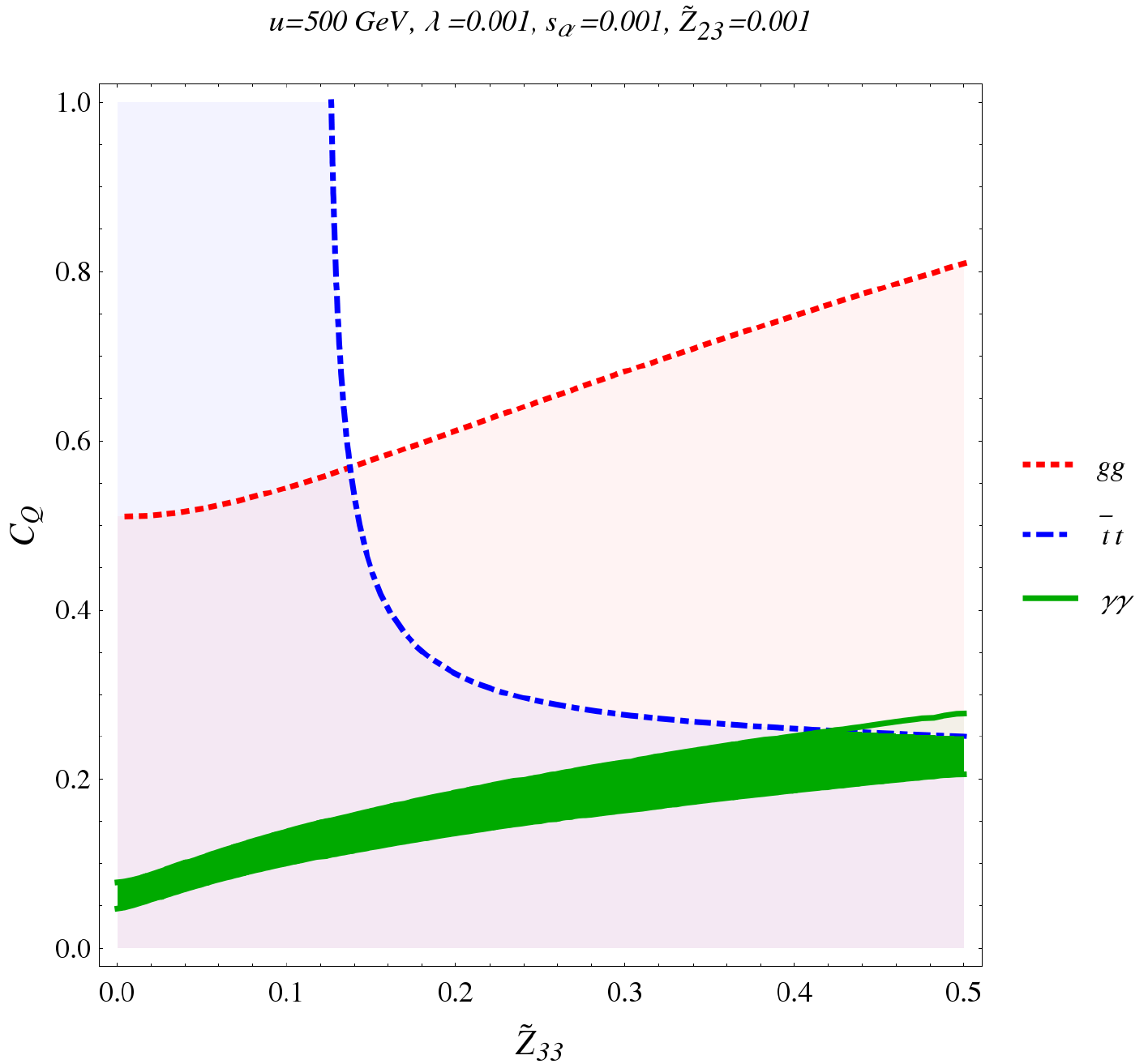}
\caption{The same as in Fig. \ref{bound1}, but for the area allowed in the $C_Q$ vs $\tilde Z_{33}$ plane.
\label{boundZ331}}
\end{figure}

We now fix the mixing angle $s_\alpha$  to the tiny value $0.001$ and find the allowed region in the $C_Q$ vs $\tilde Z_{33}$ plane. The results are shown in the left and right plots of Fig. \ref{boundZ331} for the parameter values of the scenarios of Fig. \ref{bound1}.  In this case the diboson channels $H_F\to WW$ and $H_F\to ZZ$ are considerably suppressed and  the $gg$ and $\bar{t}t$ final states are the only ones that provide useful constraints. Furthermore, in this scenario the Higgs-flavon decay width is completely dominated by the $H_F\to gg$ and $H_F\to \bar{t}t$ decay widths.

For typical values of  parameter values lying inside the allowed area the total decay width of the Higgs-flavon is of  order  a few GeV at most. This seems to be in contradiction with the ATLAS data, which point to a large decay width of about
45 GeV. However,  the CMS data hint to a narrow resonance with a decay width of a few GeVs. It is expected that these estimates change considerably once more data are available, provided that the diphoton resonance is confirmed. In such a case, a more detailed analysis of the scenario posed by the Higgs-flavon model would be in order.

\section{Conclusions and outlook}

In this work, we have studied  a model including one Higgs doublet that  participates in the spontaneous symmetry breaking and an extra inert doublet, which contains a DM candidate, together with a FN scalar field. We have found that  this model  allows for  an interesting phenomenology  to be searched for at the LHC. For instance, mixing of the Higgs
doublets with a Higgs-flavon field $H_F$ is included, which generates the Yukawa hierarchies and might induce
flavor violating Higgs couplings at non-negligible rates.
Constraints on these couplings, derived from the Higgs searches at the LHC, and their implications
for scalar anomalies  were studied. It was  found that  this model allows for a region of the parameter space where a  branching ratio of the SM Higgs decay $h \to \bar{\mu}\tau$ can be at the level of the current experimental bound.
We examined the possibility that the scalar Higgs-flavon $H_F$ could be identified with the 750 GeV scalar resonance preliminarily  observed at the LHC13 in the two-photon final state, and found the allowed area of the parameter space consistent with the ATLAS and CMS constraints on the rate of the $\sqrt{s}=8$ TeV $pp\to S\to X$ production cross section for a $750$ GeV scalar resonance decaying into weak gauge bosons, gluons, and top quark pairs. In order to reproduce the diphoton signal, the parameter space of the model must be tightly constrained,  though a tiny area  consistent with the experimental data still would survive. Furthermore, in this scenario  the total decay width of the Higgs-flavon would be of the order of a few GeVs, as preferred by the CMS data, and decays channels such as $H_F\to hh$ and $H_F\to \bar{\mu}\tau$ would be highly suppressed, with branching ratios below the $10^{-6}$ level.   Another possibility is that a  pseudoscalar Higgs-flavon $A_F$ predicted by the model could be identified with the 750 GeV resonance, but the analysis and conclusions would be rather similar. A definitive conclusion could be drawn once more data are available. We also examined  other scenario in which the Higgs-flavon is not identified with the 750 GeV resonance. In this case there is an area of the parameter space in which it is  feasible  that the decay $H_F\to hh$ could have a significant branching ratio, which would  open up the possibility for its study at the LHC. A more lengthy and detailed study of the phenomenology of this model will be published elsewhere.

 \section*{Acknowledgements}
{Support from CONACYT-SNI (Mexico) and VIEP(BUAP) are acknowledged.
}

\end{document}